# Nonlinear Model Predictive Control for Preview-Based Traction Control

Gaetano Tavolo[1], Kai Man So[1], Davide Tavernini[1], Pietro Perlo[2], Aldo Sorniotti[1*]
[1]University of Surrey, Guildford, GU2 7XH, United Kingdom
[2]Interactive Fully Electrical VehicleS (I-FEVS), La Loggia (TO), 10040, Italy
*E-mail: a.sorniotti@surrey.ac.uk

This study presents a nonlinear model predictive control (NMPC) formulation for preview-based traction control, which uses the information on the expected tire-road friction coefficient ahead to enhance the wheel slip control performance, in the context of connected vehicles with V2X features. Proof-of-concept experiments on an electric vehicle prototype highlight the real-time capability of the controller, and the wheel slip control performance improvement brought by the tire-road friction coefficient preview. Finally, an experimentally validated simulation model is used in sensitivity analyses, to evaluate the performance benefit of the preview-based controller for different dynamic characteristics (e.g., time constant and pure time delays) of the electric powertrains.

Topics: Active safety systems, chassis control, vehicle dynamics theory

## 1 INTRODUCTION

Powertrain electrification and vehicle connectivity to other road users, the infrastructure, and the cloud, usually referred to as V2X, are going to become widespread in next generation vehicles [1]-[2]. Although a wide literature describes the energy efficiency benefits of V2X connectivity for functionalities such as platooning, there is a gap in the analysis of vehicle connectivity for enhancing the performance of active safety controllers.

For example, connected vehicles could be used as moving tire-road friction sensors through the fusion of the information from cameras and optical sensors [3], or from conventional on-board state estimators [4]. Via V2X, this information would be sent to the cloud. This would determine the position of potential low tire-road friction patches, which would be communicated to the upcoming vehicles [2]. Therefore, preview-based – or pre-emptive – traction controllers can take advantage of the information on the tire-road friction ahead.

Nonlinear model predictive control (NMPC), which solves an optimization problem along a finite horizon, is the natural control choice for preview-augmented traction control, as it can embed the predicted friction coefficient profile while considering the non-linearities of the vehicle system and tires. In [5], an explicit NMPC formulation was proposed for traction control and assessed for varying tire-road friction conditions. The explicit nature reduces the computational effort required for the real-time implementation of the algorithm, at the price of increased memory utilization. In [6], implicit real-time NMPC implementations with tire-road friction preview from V2X were developed for traction control and anti-jerk control.

This paper expands upon [6], and focuses on: i) proof-of-concept experiments with the proposed traction controller with tire-road friction preview, applied to an electric vehicle (EV) prototype with an on-board powertrain; and ii) a simulation-based sensitivity analysis on the performance benefit of the proposed preview-based traction controller for vehicles with different powertrain characteristics, such as time constants and/or pure time delays, in response to a torque request variation.

## 2 CASE STUDY VEHICLE

The EV prototype developed within the European Horizon 2020 projects TELL and Multi-Moby (Fig. 1) is used as a case study in this research. The EV has a front centralized on-board electric powertrain, which is connected to the wheels through a single-speed transmission, an open differential, half-shafts, and constant velocity joints.

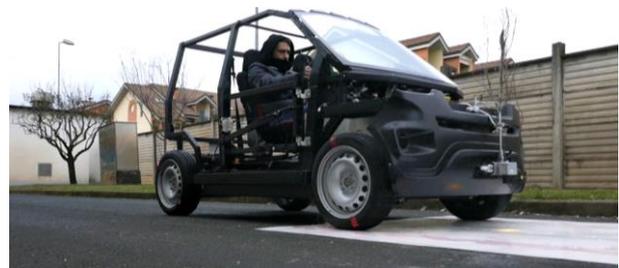

Fig. 1 The case study EV prototype during a traction control test, with a step change from high (dry tarmac) to low (white boards covered with water and soap) tire-road friction coefficient.

The EV is equipped with: i) a set of vehicle dynamics sensors, e.g., to measure the individual wheel speeds and longitudinal and lateral velocities; ii) an integrated PCAN GPS (global positioning system) with an inertial measurement unit (IMU); and iii) a dSPACE

MicroAutoBox II system for the rapid control prototyping of the traction controller.

## 3 THE CONSIDERED CONTROLLERS

Two traction controllers were developed: i) a pre-emptive NMPC implementation with preview of the tire-road friction ahead in the context of V2X; and ii) a benchmarking non-pre-emptive NMPC, which is only aware of the instantaneous tire-road friction condition, which is kept constant along the prediction horizon.

### 3.1 Control architecture

Fig. 2 is a simplified schematic of the pre-emptive NMPC architecture, where the traction controller modifies the torque requested by the human or automated driver to ensure appropriate wheel slip levels. The benchmarking controller has the same NMPC structure, except that it only knows the current tire-road friction condition. The controllers were implemented in the Matlab-Simulink environment through the ACADO toolkit [7].

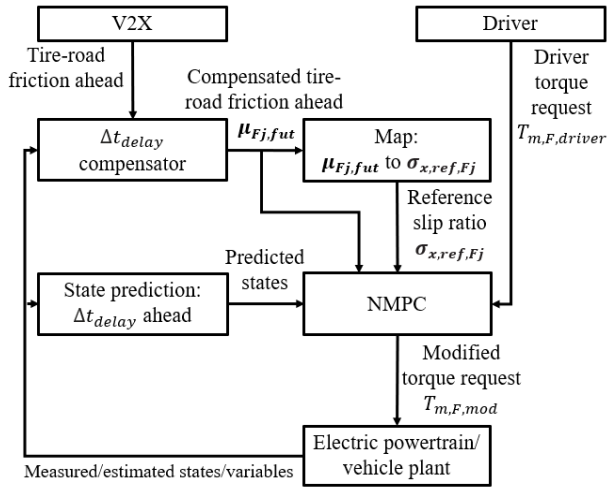

Fig. 2 Simplified block diagram of the pre-emptive traction control architecture.

### 3.2 Internal model

The traction controllers were defined for an EV with a front centralized onboard motor, but can be easily modified for other configurations. The internal model is similar to the one defined in [5] and [6], expressed through the following continuous time formulation:
$$\dot{x}(t) = f(x(t), u(t)) \quad (1)$$
where the state vector $x$ is:
$$x = [T_{m,F}, \omega_{FL}, \omega_{FR}, s_{FL}, s_{FR}, e_{int,FL}, e_{int,FR}] \quad (2)$$
where $T_{m,F}$ is the front motor torque; $\omega_{Fj}$ are the angular wheel speeds, with the subscript $F$ indicating the front axle, and $j = L, R$ indicating the left or right sides of the vehicle; $s_{Fj}$ are the longitudinal wheel slip speeds; and $e_{int,Fj}$ are the time integrals of the wheel slip ratio error.

The control action is defined as:
$$u = [T_{m,F,mod}, \varepsilon_{\sigma_x,FL}, \varepsilon_{\sigma_x,FR}] \quad (3)$$
where $T_{m,F,mod}$ is the modified front motor torque; and $\varepsilon_{\sigma_x,Fj}$ are the slack variables of the longitudinal tire slip ratios, which allow implementation of soft constraints.

To enable accurate pre-emptive torque control, the first equation of the internal model considers the powertrain dynamics:
$$\dot{T}_{m,F} = \frac{T_{m,F,mod} - T_{m,F}}{\tau} \quad (4)$$
where $\tau$ is the time constant of the electric powertrain.

The second and third equations (one each for $j = L, R$) describe the angular wheel speed dynamics:
$$\dot{\omega}_{Fj} = \frac{T_{wh,Fj} - F_{x,Fj} R_F}{J_{\omega,F}} \quad (5)$$
(5) results from the wheel moment balance, where $J_{\omega,F}$ is the front wheel mass moment of inertia. $T_{wh,Fj}$ is the individual front wheel torque, which consists of the motor torque contribution, equally divided between the two front wheels (open differential configuration):
$$T_{wh,Fj} = \frac{T_{m,F} G_{r,F}}{2} \quad (6)$$
where $G_{r,F}$ is the transmission gear ratio.

The second term on the right-hand side of (5) consists of the product of the front tire radius, $R_F$, and the longitudinal tire force, $F_{x,Fj}$, which is computed through a simplified version of the Pacejka magic formula (MF) [8]:
$$F_{x,Fj} = \mu_{x,Fj} F_{z,Fj} \quad (7)$$
where $F_{z,Fj}$ is the vertical tire load, considered constant along the prediction horizon; and $\mu_{x,Fj}$ is the longitudinal tire force coefficient:
$$\mu_{x,Fj} = D_{Fj} \sin(C_{0,F} \tan^{-1}(B_{Fj} \sigma_{x,Fj})) \quad (8)$$
with $\sigma_{x,Fj}$ being the longitudinal tire slip ratio:
$$\sigma_{x,Fj} = \frac{s_{Fj}}{\omega_{Fj} R_F} \quad (9)$$
and the longitudinal wheel slip speed, $s_{Fj}$, being approximated as:
$$s_{Fj} = \omega_{Fj} R_F - V \quad (10)$$
where $V$ is the longitudinal vehicle speed.

$B_{Fj}$ and $D_{Fj}$ are the MF parameters:
$$B_{Fj} = B_{0,F} / \mu_{Fj,fut} \quad (11)$$
$$D_{Fj} = D_{0,F} \mu_{Fj,fut} \quad (12)$$
which have been scaled based on the tire-road friction coefficient $\mu_{Fj,fut}$ obtained from V2X ($\mu_{Fj,fut}$ will be elaborated in (19)). $B_{0,F}$, $C_{0,F}$ and $D_{0,F}$ are the nominal MF front tire parameters for high friction conditions (identical on the left and right tires).

The slip dynamics (fourth and fifth model equations, one for each vehicle side) are obtained from the time derivative of (10):
$$\dot{s}_{Fj} = \dot{\omega}_{Fj} R_F - \dot{V} \quad (13)$$
with $\dot{\omega}_{Fj}$ being defined in (5), while $\dot{V}$ is obtained from the simplified longitudinal force balance equation:
$$\dot{V} = \frac{F_{x,FL} + F_{x,FR}}{m} \approx \frac{2 F_{x,Fj}}{m} \quad (14)$$
which assumes that each front wheel equally contributes to the vehicle acceleration, with $m$ being the vehicle mass.

Substituting (5) and (14) into (13), the longitudinal wheel slip speed equation becomes:

$$\dot{s}_{Fj} = \left[-\frac{R_F^2}{J_{\omega,F}} - \frac{2}{m}\right]F_{x,Fj} + \frac{T_{wh,Fj}R_F}{J_{\omega,F}} \quad (15)$$

The last two equations of the internal model consider the integral of the wheel slip ratio error, $e_{int,Fj}$, for each vehicle side:

$$\dot{e}_{int,Fj} = e_{Fj} \quad (16)$$

where the error $e_{Fj}$ is defined as the difference between the reference tire slip ratio, $\sigma_{x,ref,Fj}$, and the actual ratio, $\sigma_{x,Fj}$:

$$e_{Fj} = \sigma_{x,ref,Fj} - \sigma_{x,Fj} \quad (17)$$

$\sigma_{x,ref,Fj}$ is a function of the estimated vertical tire load and tire-road friction condition $\mu_{Fj,fut}$:

$$\sigma_{x,ref,Fj} = f_{\sigma_{x,ref,F}}(\mu_{Fj,fut}, F_{z,Fj}) \quad (18)$$

The pre-emptive capability of the controller derives from $\boldsymbol{\mu}_{Fj,fut}(k)$ at the current time step $k$ being a vector (in this manuscript, vectors are indicated in bold), obtained from V2X. $\boldsymbol{\mu}_{Fj,fut}(k)$ is computed from $\boldsymbol{S}_{fut}(k)$, which is the vector of the expected future traveled distance values along the prediction horizon:

$$\boldsymbol{\mu}_{Fj,fut}(k) = f_{\mu_{Fj,fut}}[\boldsymbol{S}_{fut}(k) + \Delta x_{delay}(k)\boldsymbol{1}] \quad (19)$$

where in the specific implementation $f_{\mu_{Fj,fut}}$ is set as a map. For computational efficiency, $\boldsymbol{S}_{fut}(k)$ is generated under the constant speed $V(k)$ assumption in the look-ahead period:

$$\boldsymbol{S}_{fut}(k) = S(k)\boldsymbol{1} + V(k)[\boldsymbol{t}_{fut} - t_{fut,0}\boldsymbol{1}] \quad (20)$$

where $S(k)$ is the current vehicle position, and $\boldsymbol{1}$ is an all-ones vector with dimension $N+1$, with $N$ being the number of steps of the prediction horizon. $\boldsymbol{t}_{fut} = [t_{fut,0}, t_{fut,1}, ..., t_{fut,N}]$ is the vector of future time values, defined for $N$ points evenly spaced according to a constant time step $T_s$, where $t_{fut,0}$ is the current time instant.

A novel feature of the pre-emptive controller is the delay compensation algorithm to account for pure time delays, $\Delta t_{delay}$, between the driver torque request variation and the powertrain response. The delay compensator advances the map of tire-road friction coefficient by a distance $\Delta x_{delay}$ under a constant speed assumption:

$$\Delta x_{delay}(k) = V(k)\Delta t_{delay} \quad (21)$$

Simultaneously, similarly to [5], the state vector $\boldsymbol{x}$ at $\Delta t_{delay}$ ahead is predicted externally to the NMPC, by using the same internal model equations. The resulting $\boldsymbol{x}$ is then used as the current (i.e., initial) state vector input $\boldsymbol{x}_{in}$ for the NMPC (see Fig. 2).

In contrast, for the non-pre-emptive NMPC implementation, $\boldsymbol{\mu}_{Fj,fut}$ is a vector of identical components, equal to the current tire-road friction condition, and $\Delta t_{delay} = \Delta x_{delay} = 0$.

### 3.3 Nonlinear optimal control problem

A generic nonlinear optimal control problem can be defined as the minimization of the following cost function in discrete time form:

$$\min_{\boldsymbol{u}} J(\boldsymbol{x}(0), \boldsymbol{u}(\cdot)) := \sum_{n=0}^{N-1} l(\boldsymbol{x}_n, \boldsymbol{u}_n) \quad (22)$$

s.t.
$$\begin{aligned}
\boldsymbol{x}_0 &= \boldsymbol{x}_{in}(k) \\
\boldsymbol{x}_{n+1} &= f_d(\boldsymbol{x}_n, \boldsymbol{u}_n) \\
\underline{\boldsymbol{x}} &\leq \boldsymbol{x}_n \leq \overline{\boldsymbol{x}} \\
\underline{\boldsymbol{x}} &\leq \boldsymbol{x}_N \leq \overline{\boldsymbol{x}} \\
\underline{\boldsymbol{u}} &\leq \boldsymbol{u}_n \leq \overline{\boldsymbol{u}} \\
\boldsymbol{u}(\cdot) &: [0, N-1]
\end{aligned}$$

where $J$ is the cost function; $\boldsymbol{u}(\cdot)$ indicates the control sequence; $\boldsymbol{x}_{in}$ is the initial value of the state vector at the current time step $k$, obtained from the available sensor measurements and state predictors; $N$ defines the prediction horizon $H_P = N T_s$; $\underline{\boldsymbol{x}}$ and $\overline{\boldsymbol{x}}$ are the lower and upper limits for $\boldsymbol{x}$; $\underline{\boldsymbol{u}}$ and $\overline{\boldsymbol{u}}$ are the lower and upper limits for $\boldsymbol{u}$; $\boldsymbol{x}_{n+1} = f_d(\boldsymbol{x}_n, \boldsymbol{u}_n)$ is the discretized version of (1); and $l(\boldsymbol{x}_n, \boldsymbol{u}_n)$ is the stage cost function associated with each time step, which is defined as:

$$\begin{aligned}
l(\boldsymbol{x}_n, \boldsymbol{u}_n) = &W_{u,\varepsilon_{\sigma_{x,FL}}}\varepsilon_{\sigma_{x,FL},n}^2 + W_{u,\varepsilon_{\sigma_{x,FR}}}\varepsilon_{\sigma_{x,FR},n}^2 \\
&+ W_{u,T_{m,F}}[T_{m,F,driver} - T_{m,F,mod,n}]^2 \quad (23) \\
&+ W_{u,e_{int,FL}}e_{int,FL,n}^2 + W_{u,e_{int,FR}}e_{int,FR,n}^2
\end{aligned}$$

where $T_{m,F,driver}$ is the motor torque request by the driver, considering only traction conditions, and assumed constant along the prediction horizon. $W_{u,\varepsilon_{\sigma_{x,Fj}}}$, $W_{u,T_{m,F}}$, and $W_{u,e_{int,Fj}}$ are the cost function weights for penalizing the slack variables, control action, and integral error. The subscript $n$ indicates the position of the step along the prediction horizon.

The constraints are:
$$\begin{aligned}
0 &\leq T_{m,F,mod,n} \leq T_{m,F,driver} \\
e_{Fj,n} + \varepsilon_{\sigma_{x,Fj},n} &\geq 0 \quad (24) \\
\varepsilon_{\sigma_{x,Fj},n} &\geq 0
\end{aligned}$$

The first condition of (24) is a hard constraint, and states that the modified torque request in Fig. 2 can only be a reduction of the driver torque demand, and not an increase. The remaining conditions refer to a soft constraint of the longitudinal wheel slip ratio error.

The controller maintains $T_{m,F,mod} = T_{m,F,driver}$ unless it predicts, in the case of the pre-emptive controller (or detects in the case of the non-pre-emptive controller) that $e_{Fj,n}$ will become (or currently is) negative, i.e., $\sigma_{x,ref,Fj} < \sigma_{x,Fj}$. In this scenario, the controller will generate an output $T_{m,F,mod} < T_{m,F,driver}$ to reduce the torque. The integral terms in (23), weighted by $W_{u,e_{int,Fj}}$, tend to compensate for external disturbances or internal model mismatches, by reducing the steady-state errors.

## 4 VEHICLE EXPERIMENTS
### 4.1 Experiment setup

Proof-of-concept experiments were performed on the case study EV prototype. A scenario with a step change in tire-road friction coefficient was created, where the friction coefficient is initially high, corresponding to dry tarmac conditions, which is followed by a sudden drop to a low level, corresponding to smooth plastic boards covered with water and soap, as in Fig. 1.

Three configurations were tested: i) the pre-emptive NMPC with tire-road friction preview; ii) the benchmarking non-pre-emptive NMPC; and iii) the passive configuration without controller intervention. Table 1 reports the settings for the NMPC implementations, with the pre-emptive and non-pre-emptive cases configured identically.

The powertrain time constant $\tau$ was determined through tip-in (step change from 0% to 100% torque request) and tip-out (step change from 100% to 0% torque request) tests. By averaging the $\tau$ values from both tests, it was determined a $\tau$ value of 140 ms for the case study EV. The controllers were loaded into the dSPACE MicroAutoBox II system for real-time implementation.

For the experiments, a map of the tire-road $\mu$ condition was programmed in spatial coordinates a priori. As the test distance was short, the EV position was identified through vehicle speed integration (GPS has also been tested for this purpose as in [6]). In the tests, the driver demands the maximum motor torque from standstill conditions. The vehicle initially accelerates on the high friction section. The transition to low friction for the front powered wheels occurs at ~0.29 s and ~10 km/h, see Fig. 3.

Table 1 NMPC controller settings for experiments (Exp.) and simulations (Sim.).

| Parameter | Exp. | Sim. |
|---|---|---|
| Prediction horizon $H_P$ (ms) | 250 | 250 |
| Time step $T_s$ (ms) | 25 | 5 |
| No. of steps $N$ | 10 | 50 |
| Integration time step (ms) | 1 | 1 |

**4.2    Experiment results**

The first subplot of Fig. 3 shows the motor torque profiles. The passive vehicle is not subject to torque reduction and follows the driver request. The torque reduces in the right half of the subplot because the motor speed increases, based on the motor torque characteristic as a function of speed. The non-pre-emptive NMPC configuration (labeled 'NMPC') imposes a torque reduction immediately after the vehicle enters the low friction section (see the 'NMPC $T_{m,F,mod}$' line). The pre-emptive NMPC asks for a torque reduction from ~0.1 s (see the 'Pre-NMPC $T_{m,F,mod}$' line), which is before the beginning of the low friction section, as it is pre-empting the drop in the friction coefficient that is going to occur in the near future. The actual motor torque profiles (solid lines) trail the torque requests (corresponding dashed lines) according to the powertrain dynamics.

The vehicle response is significantly different in the three configurations. In the second subplot of Fig. 3, the longitudinal vehicle speed $V$ ranges from 10 to 15 km/h in the low friction section for all three cases. However, for the passive case, the tangential speed of the front right (FR) wheel (i.e., $\omega_{FR}R_F$) hits 80 km/h, indicating significant wheel spinning. In contrast, for the non-pre-emptive NMPC, the wheel speed reaches 35 km/h, while for Pre-NMPC the wheel speed stays close to the longitudinal vehicle speed, indicating negligible wheel spinning. This is confirmed by the third subplot, which shows the front right tire slip ratio. The passive configuration has the highest peak value, hitting 0.85, followed by the non-pre-emptive NMPC, peaking at 0.70. Pre-NMPC has the best performance, with a maximum slip ratio of <0.05.

In summary, the NMPC with tire-road friction preview can significantly improve the wheel slip control performance with respect to the benchmarking NMPC without preview and the passive case.

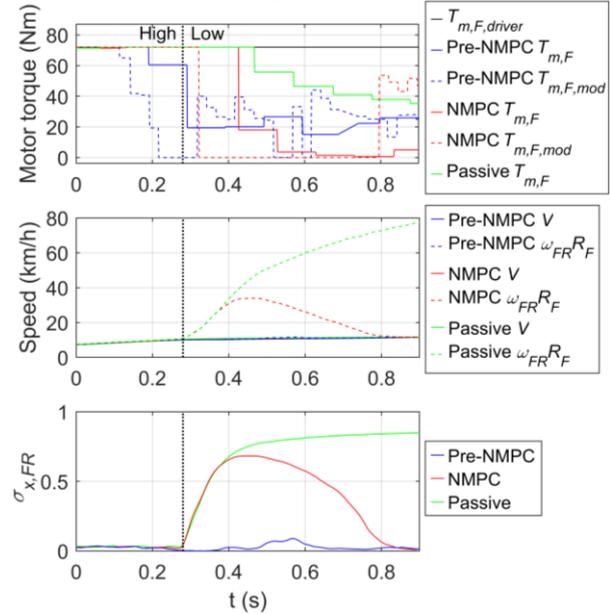

Fig. 3 Experimental traction control test results for the pre-emptive controller (Pre-NMPC), non-pre-emptive controller (NMPC) and passive vehicle configurations. The plots show the profiles of the front reference motor torque modified by the controller, $T_{m,F,mod}$, actual torque, $T_{m,F}$, and driver torque request, $T_{m,F,driver}$; the longitudinal vehicle speed, $V$, and front right (FR) tangential wheel speed, $\omega_{FR}R_F$; and the FR tire slip ratio, $\sigma_{x,FR}$. The vertical dotted line separates the high and low tire-road friction sections.

**5    SIMULATIONS**
**5.1    Simulation setup**

Simulation-based sensitivity analyses were conducted to evaluate the performance of the preview-based NMPC for different characteristics of the electric powertrain, in terms of: i) time constant $\tau$ of the torque response transfer function; and ii) pure time delay $\Delta t_{delay}$ from the torque variation request to the corresponding torque response.

A simulation model of the case study EV was implemented in the Matlab-Simulink/IPG CarMaker environment, and used to replicate the traction control experiment scenario in Section 4, with a step transition of $\mu$ from high to low levels. The model was experimentally validated in terms of vehicle and wheel dynamics (Fig. 4).

The NMPC settings for the simulations are shown in Table 1, where the number of steps has been increased with respect to the experiments. This improves the NMPC performance, under the assumption of less restrictive hardware limitations.

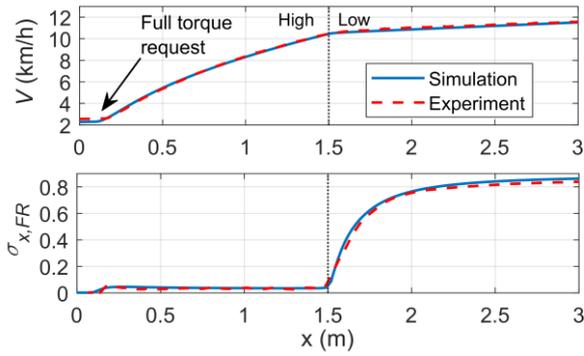

Fig. 4 Simulation model validation against experimental results for the test scenario in Fig. 1, with a step change from high to low friction. The plots show the distance profiles of $V$ and $\sigma_{x,FR}$, with the vertical dotted line separating the high and low tire-road friction sections.

### 5.2 Effect of the powertrain time constant

For the first set of simulations, $\tau$ was varied in both the internal model and the simulation model to represent the performance of different powertrains, while the time delay was set to $\Delta t_{delay} = 0$. Fig. 5 shows the performance comparison of the benchmarking NMPC without preview (left subplots), and the Pre-NMPC (right subplots).

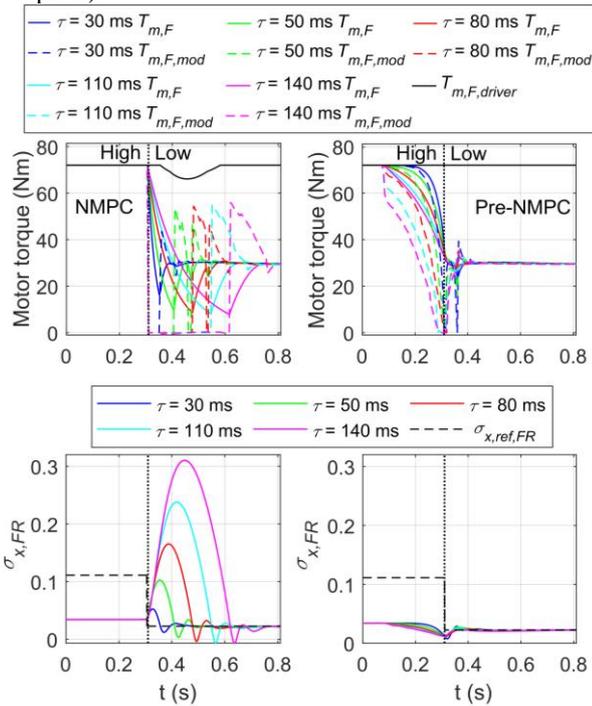

Fig. 5 Simulation results for the non-pre-emptive NMPC (left) and pre-emptive NMPC (right) configurations, applied to powertrains without pure time delay, but with different values of the powertrain time constants $\tau$. The plots show the profiles of $T_{m,F,mod}$, $T_{m,F}$, $T_{m,F,driver}$ and $\sigma_{x,FR}$. The vertical dotted lines separate the high and low tire-road friction sections.

As the benchmarking NMPC is only aware of the instantaneous $\mu$ value, the controller only instructs a torque reduction immediately after entering the low $\mu$ section at ~0.3 s, and the initial $T_{m,F,mod}$ responses are the same for the different time constant values. Therefore, the actual motor torque $T_{m,F}$ and the resultant slip ratio $\sigma_{x,FR}$ are heavily affected by the value of $\tau$. From Fig. 5, $\tau = 140$ ms, corresponding to the slowest among the considered powertrains, brings the highest peak value of $\sigma_{x,FR}$.

In contrast, the pre-emptive NMPC can effectively compensate for the powertrain dynamics by matching the time constant value of its internal model in (4) with the one of the actual powertrain $\tau$. In Fig. 5, the $\sigma_{x,FR}$ profiles for Pre-NMPC are similar and close to the reference, regardless of $\tau$. This can be explained through the motor torque subplot, where, for example, the controller for the $\tau = 140$ ms configuration takes the earliest and strongest action, by reducing $T_{m,F,mod}$ at ~0.1s. In the configurations with smaller $\tau$, the controller acts later with a gentler torque reduction, and still achieves a similarly desirable $\sigma_{x,FR}$ profile due to the faster powertrain response.

### 5.3 Effect of powertrain pure time delay

The second set of simulations studies the benefit of the pure time delay compensator in (19)-(21). $\Delta t_{delay}$ was varied while the time constant was set to $\tau = 140$ ms. Fig. 6 shows the simulation results for the Pre-NMPC configuration (the NMPC without preview is inherently unable to compensate the delay, and is therefore not presented here).

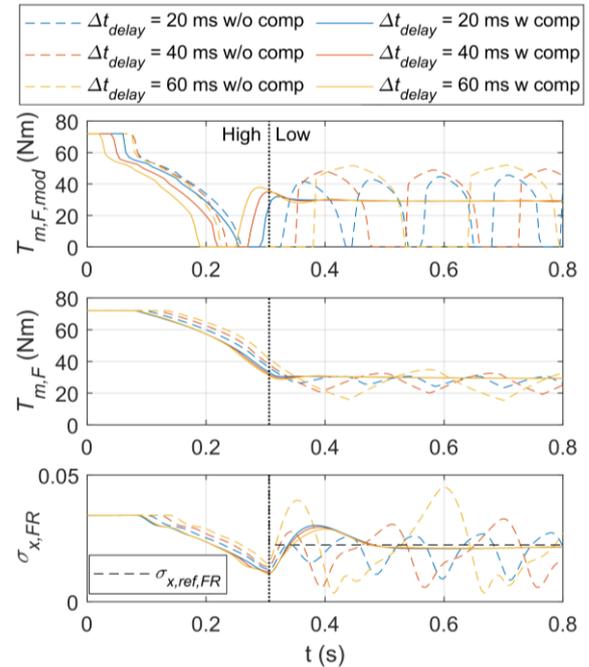

Fig. 6 Simulation results for the pre-emptive NMPC configuration, applied to powertrains with $\tau = 140$ ms and different values of $\Delta t_{delay}$, with ('w') and without ('w/o') the proposed delay compensation algorithm. The plots show the profiles of $T_{m,F,mod}$, $T_{m,F}$, and $\sigma_{x,FR}$. The vertical dotted lines separate the high and low tire-road friction sections.

The dashed lines in the figure represent the scenario without delay compensation, i.e. a non-zero $\Delta t_{delay}$ is present in the powertrain, however, $\Delta t_{delay}$ is set to 0 in (21), and thus the controller is unaware of the powertrain delay. The largest tested delay was $\Delta t_{delay}$ = 60 ms, which brings the worst performance among the three cases without compensation, with significant torque and tire slip ratio oscillations.

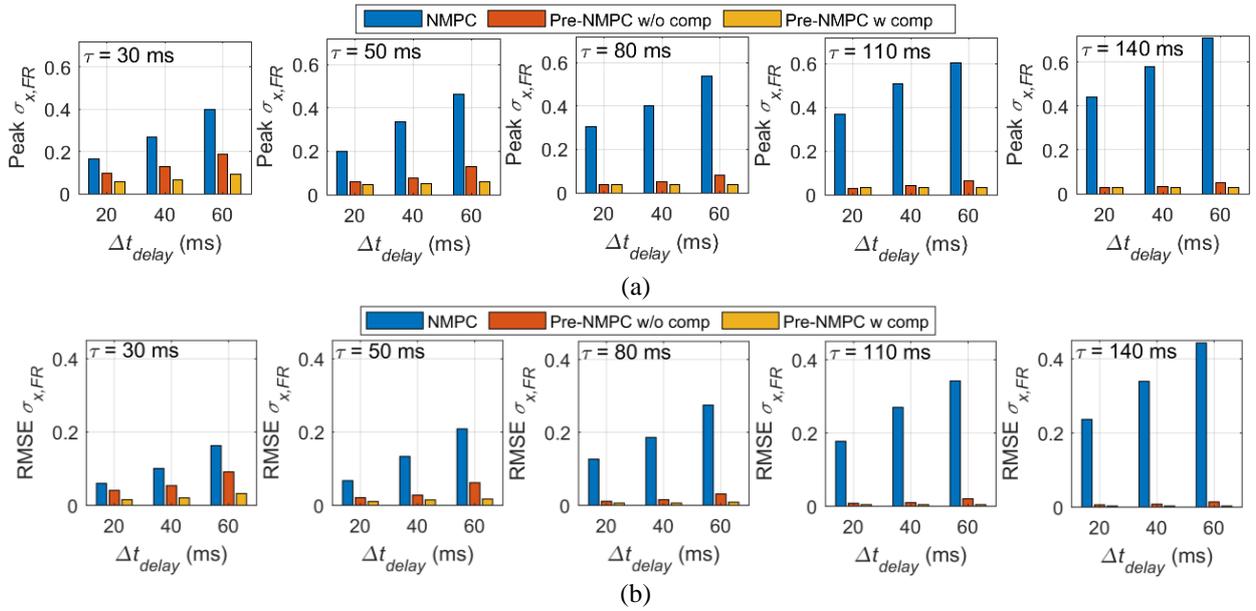

Fig. 7 Simulation results in terms of peak and RMSE values of $\sigma_{x,FR}$, for different values of $\tau$ and $\Delta t_{delay}$, for the non-pre-emptive NMPC ('NMPC' in the legend), pre-emptive NMPC without $\Delta t_{delay}$ compensation (Pre-NMPC w/o comp) and pre-emptive NMPC with $\Delta t_{delay}$ compensation (Pre-NMPC w comp).

In contrast, the three configurations with delay compensation have similar $T_{m,F}$ and $\sigma_{x,FR}$ profiles, with $\sigma_{x,FR}$ values that are very close to the reference. Similarly to what was observed in Section 5.2, the configuration with the largest delay, $\Delta t_{delay}$ = 60 ms, takes the earliest pre-emptive action to compensate for the long delay.

### 5.4   Overall trends

Further simulations were performed for NMPC and Pre-NMPC, the latter with and without delay compensation, for different combinations of $\tau$ and $\Delta t_{delay}$, to observe the resulting trends. The controller cost function weights were the same for all configurations. Fig. 7 is a summary of the results, in terms of peak value of $\sigma_{x,FR}$, and root mean square value of the slip ratio error $\sigma_{x,ref,FR} - \sigma_{x,FR}$ (RMSE $\sigma_{x,FR}$ in the figure), computed along the simulated tests.

For the benchmarking NMPC configuration, larger $\tau$ and/or $\Delta t_{delay}$ correspond to larger peak and RMSE values of $\sigma_{x,FR}$. The pre-emptive NMPC without $\Delta t_{delay}$ compensation brings a significant improvement compared to the benchmarking NMPC, although its performance still worsens with increasing $\Delta t_{delay}$. In contrast, the pre-emptive NMPC with compensation achieves consistent performance regardless of $\Delta t_{delay}$.

### 6   CONCLUSION

The main conclusions are: i) proof-of-concept experiments on an EV prototype show that the proposed traction controller with tire-road friction preview can run in real-time, and pre-emptively reduce wheel torque and tire slip in maneuvers with abrupt friction coefficient reductions; and ii) it is possible for the preview controller, differently from the benchmarking controller, to effectively compensate for a wide range of dynamic characteristics of the electric powertrains.

### FUNDING

This work was supported by the European Union's Horizon 2020 programme under grant agreements no. 824254 (TELL) and no. 101006953 (Multi-Moby).

### REFERENCES

[1] N. Lu, N. Cheng, N. Zhang, X. Shen, and J.W. Mark, "Connected Vehicles: Solutions and Challenges," *IEEE Internet of Things Journal,* vol. 1, no. 4, pp. 289-299, 2014.

[2] U. Montanaro, S. Dixit, S. Fallah, M. Dianati, A. Stevens, D. Oxtoby, and A. Mouzakitis, "Towards connected autonomous driving: review of use-cases," *Vehicle System Dynamics,* vol. 57, no. 6, p. 779–814, 2019.

[3] S. Khaleghian, A. Emami, and S. Taheri, "A technical survey on tire-road friction estimation," *Friction,* vol. 5, no. 2, pp. 123-146, 2017.

[4] S. Antonov, A. Fehn, and A. Kugi, "Unscented Kalman filter for vehicle state estimation," *Vehicle System Dynamics,* vol. 49, no. 9, pp. 1497-1520, 2011.

[5] D. Tavernini, M. Metzler, P. Gruber, and A. Sorniotti, "Explicit Nonlinear Model Predictive Control for Electric Vehicle Traction Control," *IEEE Transactions on Control Systems Technology,* vol. 27, no. 4, pp. 1438-1451, 2019.

[6] A. Scamarcio, C. Caponio, M. Mihalkov, P. Georgiev, J. Ahmadi, K.M. So, D. Tavernini, and A. Sorniotti, "Predictive anti-jerk and traction control for V2X connected electric vehicles with central motor and open differential," *IEEE Transactions on Vehicular Technology,* 2022 (in press).

[7] B. Houska, H.J. Ferreau, and M. Diehl, "An auto-generated real-time iteration algorithm for nonlinear MPC in the microsecond range," *Automatica,* vol. 47, no. 10, pp. 2279-2285, 2011.

[8] H.B. Pacejka, Tire and Vehicle Dynamics, Oxford: Butterworth-Heinemann, 2012.